# Magnetotransport and magnetodielectric effect in Bi-based perovskite manganites

Asish K. Kundu,[*] R. Ranjith, V. Pralong, V. Caignaert, and B. Raveau

*CRISMAT Laboratory, ENSICAEN UMR6508, 6 Boulevard Maréchal Juin, Cedex 4,*

*Caen-14050, France*

**Abstract**

The effect of cobalt and nickel substitutions for manganese on the physical properties of the perovskite manganite $La_{1.2}Bi_{0.8}Mn_{2-x}(Ni/Co)_xO_{6\pm\delta}$, with $0.0 \leq x \leq 0.8$, has been investigated. It is observed that the ferromagnetism is enhanced, $T_C$ being increased from 103 K for the parent compound ($x = 0.0$) to 178 K for Ni phase, and to 181 K for the Co phase ($x = 0.8$). Moreover, the systems remain insulating and depict relatively large values of magnetoresistance effect at low temperatures (up to 67 % at 90K and ±70 kOe, for $x = 0.0$ phase). These phenomena are interpreted by means of electronic phase separation, where the ferromagnetic $Mn^{4+}/Ni^{2+}$ and $Mn^{4+}/Co^{2+}$ interactions reinforce the $Mn^{3+}/Mn^{4+}$ interactions by super-exchange interaction. The dielectric measurements below the magnetic transition temperatures exhibit weak magneto-dielectric effect of around 0.25% at 80K, which may be due to spin-lattice interaction.

---

*Corresponding author email: asish.kundu@ensicaen.fr/asish.k@gmail.com

## I. INTRODUCTION

The discoveries of giant/colossal magnetoresistance and multiferroic properties in manganite perovskite have been of a great interest due to their novel device applications as well as their unusual fascinating properties.[1-6] A large number of reports have already described the structural, magnetic and electron transport properties of the perovskite oxide, $LnMnO_3$ (Ln = Rare Earth or Bi).[3-5] In the last few years, there has been an increasing interest in the understanding of the basic physics/chemistry of these materials. In such systems, the key feature is the charge transfer induced by either hole/electron doping or stoichiometry defects like cations vacancy.[1-3] Such an example is the study of colossal magnetoresistance properties in the doped $La_{1-x}A_xMnO_3$ (A = alkaline earth) that began over a decade ago.[7,8]

In contrast to the trends emphasizing by $RE_{1-x}A_xMnO_3$, the role of divalent cations substitution at the bismuth site, $Bi_{1-x}A_xMnO_3$ has been reported to enhance the charge/orbital ordering.[9,10] But, no such ferromagnetism and/or metal-insulator transitions have been observed similar to doped $REMnO_3$. Although the parent $BiMnO_3$ is an insulating ferromagnet with a Curie temperature of ~105K and the insulating state is very robust.[10-12] It shows very unique phenomena like multiferroicity,[12] magnetocapacitance[13] effects at low temperature, making this compound very special in comparison to other manganite perovskites. To synthesize the multiferroic $BiMnO_3$ phase the high pressure/temperature is required[12] or can be prepared in the thin film form.[14] The coexistence of ferromagnetism and ferroelectricity as reported by many authors, is due to the orbital ordering of the $Mn^{3+}$ ions (different from others manganite perovskite); ferroelectricity is linked to the lone pair electrons of $Bi^{3+}$ ion which induces structural distortions.[3-5,11-14] It seems that the properties were well understood in terms of the numerous works devoted to this system. Nevertheless, very recently few groups[15] have reported centrosymmetry (C2/c) structure unlike that of the reported noncentrosymmetry (C2) of $BiMnO_3$. The unusual multiferroicity of $BiMnO_3$ may remain a problem of study for material scientists and therefore is subjected to further discussion.

The demands for multiferroic materials are extensive in terms of potential applications as well as in condensed matter physics to unravel the fundamental mechanism.[3-6] Despite the numerous investigations on multiferroic materials in the past few years very few perovskite-manganites have been known as ferroelectric-ferromagnet. In this respect, the perovskite $La_{1-x}Bi_xMnO_{3+\delta}$ studied for their FM properties[16,17] and recently discovered multiferroics $La_{0.1}Bi_{0.9}MnO_3$ (Ref. 18) and $La_{0.2}Bi_{0.8}MnO_3$ (Ref. 19) are of great interest. These La-substituted $BiMnO_3$ phases show multiferroic properties also at low temperature ($T \approx 100K$) similar to the parent compound. It is noticed that for $x \geq 0.6$, high-pressure/temperature synthesis is required and the multiferroicity is reported for $x = 0.8$ and 0.9 phases only in the epitaxial thin films.[16-19] The $La_{0.1}Bi_{0.9}MnO_3$ ($La_{0.2}Bi_{0.8}MnO_3$) exhibits ferroelectricity at around 300K (150K) and ferromagnetism around ~95K (90K), but for the bulk phase there is no such report.[18,19] Ferroelectricity is described as the presence of $Bi^{3+}$ lone pair electrons whereas ferromagnetism is explained by the presence of $Mn^{4+}$ ions due to cationic vacancies (i.e. volatility of Bi or nonstoichiometry phase like $LaMnO_{3+\delta}$).[16-18]

In the present investigations we have been exploring the kind of Bi-based manganite perovskites, which show simultaneously ferromagnetic and insulating properties. Actually, the ferromagnetic semiconductors/insulators are novel singular materials that could exhibit simultaneously electronic and magnetic ordering.[6] Such materials have practical application in spintronic and magnetodielectric based devices.[5,6] Also, the recent investigations on multiferroic and/or spin filtering in this type of thin films have enhanced the possibility of device applications.[18,19] We have also recently reported Bi-based manganite, which can be prepared in bulk phase at normal pressure and in the case of Co doped phase (both bulk and thin films), the FM transition $T_C$ is higher than $BiMnO_3$.[20]

In this paper, we have studied the magnetoelectric properties of the Bi-based manganite where $Mn^{3+}$ is replaced by Ni and Co ions. Bearing in mind that the substitution of bismuth for lanthanum in $LaMnO_{3+\delta}$ increases the insulating properties but tends to decrease $T_C$,[16,17] we have selected the $La_{1.2}Bi_{0.8}Mn_{2-x}(Ni/Co)_xO_{6+\delta}$ series and varied the Ni(Co) level so that we can achieve highest ferromagnetism with insulating property. Interestingly, it is observed that the FM $T_C$ is increased from 103K for $La_{1.2}Bi_{0.8}Mn_2O_{6.20}$ to 178K for $La_{1.2}Bi_{0.8}Mn_{1.2}Ni_{0.8}O_{5.88}$ and to 181K for $La_{1.2}Bi_{0.8}Mn_{1.2}Co_{0.8}O_{6.02}$, with enhancement of electrical resistivities compared to the previously reported Bi-based manganites.[17,20] Also, below $T_C$ the magnetic field dependent dielectric studies have been performed on Co doped sample which exhibits magnetodielectric coupling at low temperatures.

## II. EXPERIMENTAL PROCEDURE

The polycrystalline samples were prepared by conventional sol gel method as described in the previous work.[20] The single phase powder samples were pressed into rectangular bars, and finally sintered at 1173K in platinum crucible for 24h and cooled rapidly to room temperature by taking out the samples from the furnace. Phase purity, chemical analysis and oxygen stoichiometry were carried out for all samples as described in Ref. 20.

The magnetization, resistivity and magnetoresistance measurements were carried out by a Quantum Design physical properties measurements system (PPMS;

SQUID). The details of the magnetization and electrical resistivity measurements procedure is mentioned in our previous work.[20] The isothermal magnetoresistance measurements have been carried out in an applied field of ±70 kOe. The capacitance measurements were carried out as a function of temperature and magnetic field by the PPMS coupled with an impedance analyzer (Agilent technologies-4284A) in the range of 10-300K. In the dielectric measurements, the frequency range was 100Hz-1MHz and the applied field was ±50 kOe. The electrodes were prepared in capacitor geometry with either side of the polished pellets painted with silver paste (Dupont).

## III. RESULTS AND DISCUSSION

**A. Structure analysis**

Figure 1 shows the X-ray powder diffraction (XRPD) patterns for three samples of the series $La_{1.2}Bi_{0.8}Mn_{2-x}(Ni/Co)_xO_{6\pm\delta}$, investigated at room temperature. The single phase samples are obtained over the doping concentrations of $0.0 \leq x \leq 0.8$, without any secondary phase. Beyond $x = 0.8$, traces of impurities, involving $Bi_2O_3$ or other phases of Bi-Mn-Ni(Co)-O appear. The Rietveld[21] refinements show that the samples $La_{1.2}Bi_{0.8}Mn_2O_{6.20}$ ($x = 0.0$), $La_{1.2}Bi_{0.8}Mn_{1.2}Ni_{0.8}O_{5.88}$ ($x_{Ni} = 0.8$) and $La_{1.2}Bi_{0.8}Mn_{1.2}Co_{0.8}O_{6.02}$ ($x_{Co} = 0.8$), can be indexed in an orthorhombic structure, with *Pnma* space group (Table 1). Note that the previously reported crystal structure for $La_{1.2}Bi_{0.8}Mn_2O_{6.20}$ is rhombohedral.[16,17] Such a difference may originate from the completely different synthesis conditions and oxygen content. Moreover, the doped samples exhibit similar features with increasing doping concentrations i.e. the lattice parameters decrease gradually with increasing the substitution level. Hence, the lattice parameters/cell volumes for $La_{1.2}Bi_{0.8}Mn_{1.2}Ni_{0.8}O_{5.88}$ and $La_{1.2}Bi_{0.8}Mn_{1.2}Co_{0.8}O_{6.02}$ samples are smaller than the parent compound as presented in Table 1, which is probably due to the difference in oxygen content.

## B. Magnetic properties

The temperature dependence zero field cooled (ZFC) and field cooled (FC) magnetization data, M(T), of $La_{1.2}Bi_{0.8}Mn_{2-x}(Ni/Co)_xO_{6\pm\delta}$, ($x$ = 0.0, 0.8) in an applied field of 1000 Oe is shown in Fig. 2(a). The parent compound $La_{1.2}Bi_{0.8}Mn_2O_{6.20}$ shows a paramagnetic (PM) to ferromagnetic (FM) transition, $T_C$, around 103K [Fig. 2(a)] similar to the earlier report.[16,17] The FM transition is calculated from the minimum position of the $dM_{FC}/dT$ versus temperature plot. The substitution of manganese by nickel/cobalt in the parent $La_{1.2}Bi_{0.8}Mn_2O_{6.20}$ compound, results in a significant increase of the FM $T_C$ with increase in the doping concentration. The highest values of FM $T_C$ ~178K and 181K are obtained for the compositions $La_{1.2}Bi_{0.8}Mn_{1.2}Ni_{0.8}O_{5.88}$ and $La_{1.2}Bi_{0.8}Mn_{1.2}Co_{0.8}O_{6.02}$ respectively [Fig. 2(a)]. In the intermediate doping concentration, the $T_C$ value is higher than that of the parent phase with relatively lower value of magnetization (not shown here). Moreover, the ZFC and FC magnetization data depict considerably large divergence for Co doped samples, whereas Ni doped samples show rather weak irreversibility at a lower magnetic field (≤ 100 Oe), akin to the parent compound. This indicates that the magnetic interactions in Ni and Co doped systems are completely different from the long range ferromagnetism. To further establish the different types of magnetic interactions we have carried out the low temperature magnetic measurements in details, which will be discussed later. In Fig. 2(b), we have shown the ZFC and FC magnetization data of $La_{1.2}Bi_{0.8}Mn_{1.2}Co_{0.8}O_{6.02}$, studied in the applied fields of 100 and 5000 Oe respectively. It shows that the irreversible temperature ($T_{irr}$) obtained from the ZFC and FC magnetization curves is $T_{irr}$ ~179 K and a sharp cusp in the ZFC data (at H = 100 Oe), which are very close to $T_C$ [Fig. 2(b)]. With increasing the field value the $T_{irr}$ shift towards lower temperature and becomes $T_{irr} < T_C$. Therefore, in an applied field of 5000 Oe the $T_{irr}$ value is ~110K and the peak in the ZFC curve broadens and shifts towards lower temperature [Fig. 2(b)]. Such a behavior is similar to that of a spin or cluster glass type material as reported in the literature.[22,23] In contrast, for the $La_{1.2}Bi_{0.8}Mn_2O_{6.20}$ and $La_{1.2}Bi_{0.8}Mn_{1.2}Ni_{0.8}O_{5.88}$ samples, with a higher field of 1000 Oe, both the ZFC and FC curves merge down to low temperatures and the peak in ZFC curve disappears completely [Fig. 2(b)].

We have also studied the high temperature region to investigate the magnetic interactions above $T_C$. In the PM region the magnetization curve is well fitted to the Curie-Weiss law. As shown in Fig. 3, the temperature dependence inverse magnetic susceptibility data in the temperature range of 200-400K, follow a linear behavior and from the fitting ($T \geq 275K$) we have calculated the PM Weiss temperature ($\theta_p$) and the effective magnetic moment ($\mu_{eff}$). The $\theta_p$ value increases from 140K for parent $La_{1.2}Bi_{0.8}Mn_2O_{6.20}$ to 200K and 205K for $La_{1.2}Bi_{0.8}Mn_{1.2}Ni_{0.8}O_{5.88}$ and $La_{1.2}Bi_{0.8}Mn_{1.2}Co_{0.8}O_{6.02}$ compounds respectively. The corresponding $\mu_{eff}$ values are ~7.7 $\mu_B$/f.u., 5.7 $\mu_B$/f.u. and 6.7 $\mu_B$/f.u. for the three compounds respectively. The obtained $\theta_p$ values are always higher than the $T_C$ values, and the positive value signifies the FM interactions in the high temperature region as well. The positive magnetic interactions in the parent compound, $La_{1.2}Bi_{0.8}Mn^{III}_{1.6}Mn^{IV}_{0.4}O_{6.20}$, is due to the presence of $Mn^{3+}$ and $Mn^{4+}$ ions, which favor the FM interaction. In the case of Ni and Co doped samples the superexchange interactions between $Mn^{4+}$-O-$Ni^{2+}(Co^{2+})$ provide the positive FM interactions according to Goodenough-Kanamori rule, as reported earlier for doped $LaBiMn_2O_{6+\delta}$ systems.[20] Hence, the FM transition temperature increases with increasing the Ni(Co) substitution level and the magnetic behavior is almost similar for all Ni(Co) doped samples.

The low temperature magnetic phase has been investigated in details to characterize the nature of FM interactions between the different magnetic ions. The isothermal magnetic hysteresis loops, $M(H)$, have been studied at different temperatures (Fig. 4). For the $La_{1.2}Bi_{0.8}Mn_{1.2}Ni_{0.8}O_{5.88}$ sample one indeed observes a very small hysteresis loop at 10K, with a remanent magnetization ($M_r$) value of ~ 0.61 $\mu_B$/f.u. and a coercive field ($H_C$) of ~ 85 Oe. These values are almost similar to the $La_{1.2}Bi_{0.8}Mn_2O_{6.20}$ compound [inset of Fig. 4(a)], which shows an irreversible magnetization value at 10K ($M_r$ ~ 0.30 $\mu_B$/f.u. and $H_C$ ~ 51 Oe). At low temperature, the relatively smaller value of $H_C$ signifies a typical soft ferromagnet and above $T_C$ the $M(H)$ behavior is linear, corresponding to a PM state. In contrast, the hysteresis loops of $La_{1.2}Bi_{0.8}Mn_{1.2}Co_{0.8}O_{6.02}$ are relatively large with a highest coercive field of 2.7 kOe at 10K similar to a hard ferromagnet [inset of Fig. 4(c)]. The highest value of magnetic moment is obtained ~ 6.8 $\mu_B$/f.u. for the parent compound, although it is less than the spin only value of Mn ions, whereas, for $La_{1.2}Bi_{0.8}Mn_{1.2}Ni_{0.8}O_{5.88}$ and $La_{1.2}Bi_{0.8}Mn_{1.2}Co_{0.8}O_{6.02}$ the corresponding values are 4.1 $\mu_B$/f.u. and 5.4 $\mu_B$/f.u. respectively. The lowest value of moment for $La_{1.2}Bi_{0.8}Mn_{1.2}Ni_{0.8}O_{5.88}$ sample may be due to its lower oxygen content which indirectly controls the ratio of $Mn^{3+}$ and $Mn^{4+}$ ions. For example in $La_{1.2}Bi_{0.8}Mn^{III}_{0.4}Mn^{IV}_{0.8}Co^{II}_{0.8}O_{6.02}$, the FM interaction between $Mn^{4+}$-O-$Co^{2+}$ dominates over $Mn^{3+}$-O-$Mn^{3+}$ AFM interaction. In contrast, in the case of $La_{1.2}Bi_{0.8}Mn^{III}_{0.64}Mn^{IV}_{0.56}Ni^{II}_{0.8}O_{5.88}$ the FM interaction between $Mn^{4+}$-O-$Ni^{2+}$ is weaker compared to $Mn^{4+}$-O-$Co^{2+}$. This is due to a reduced amount of oxygen content that leads to a higher percentage of $Mn^{3+}$ ions. Therefore, the negative AFM interactions ($Mn^{3+}$-O-$Mn^{3+}$ and $Ni^{2+}$-O-$Ni^{2+}$) are stronger than in the parent $La_{1.2}Bi_{0.8}Mn_2O_{6.20}$ and $La_{1.2}Bi_{0.8}Mn_{1.2}Co_{0.8}O_{6.02}$ phases and the overall magnetic moment is smaller for $La_{1.2}Bi_{0.8}Mn_{1.2}Ni_{0.8}O_{5.88}$ phase. Another interesting feature at low temperature is the unsaturated behavior of the *M(H)* curves even at higher fields for all samples, which is a characteristic feature of glassy ferromagnets.[20,24]

From the field dependent magnetic measurements it is confirmed that the FM and AFM components coexist at low temperatures. Indeed the data show the superposition of two types of contribution below the FM transitions. The FM component is characterized by an irreversible loop and a finite value of coercive field and the AFM component is characterized by unsaturated value of the magnetization. The latter increases almost linearly with increasing the applied field. The observed *M(H)* behavior in these compounds is due to the presence of different magnetic interactions. Therefore, at low temperatures there must be a subtle balance between the FM and AFM interactions or in other words the system is electronically phase separated into FM and AFM clusters. Hence, there will always be a competition between these two interactions to dominate one over another giving rise to a tendency of glassy FM state in the material.[20,23,24]

In order to further characterize the low temperature magnetic phase of the present systems, we have carried out frequency dependent magnetic measurements which is a very efficient way to investigate the magnetic glassy behavior. Figure 5 shows the temperature dependent in phase $\chi'(T)$ component of the ac-susceptibility measured at four different frequencies. The in phase $\chi'(T)$ data reveals similar features as the low field ZFC magnetization data. The parent compound shows a sharp maximum corresponding to FM $T_C$, which is frequency independent [Fig. 5(a)]. Similarly, $La_{1.2}Bi_{0.8}Mn_{1.2}Ni_{0.8}O_{5.88}$ exhibits an abrupt increase of $\chi'(T)$ at ~190K characteristic of FM ordering, which is almost frequency independent [Fig. 5(b)]. This behavior suggests that below $T_C$ the FM states originate from the intra cluster ferromagnetism i.e. the cluster glass type behavior rather than typical long range ferromagnetism. In contrast, the $La_{1.2}Bi_{0.8}Mn_{1.2}Co_{0.8}O_{6.02}$ compound [Fig. 5(c)] shows a clear frequency dependent peak below $T_C$ (~181K). The out of phase $\chi''(T)$ component depicts similar features around the same temperature. This is also related to intra cluster FM interactions which give rise to a spin glass like behavior at low temperature. The broad peak in $\chi'(T)$, occurs around 179K at 10 Hz in an ac field of 10 Oe, which shifts toward higher temperature (180K) with increasing frequency (10kHz). This is a characteristic feature of the canonical spin glass system[22] and previously reported for $LaBiMn_{4/3}Co_{2/3}O_{6.02}$ phase.[20] Hence, from the present ac magnetic study it is confirmed that at low temperature the $La_{1.2}Bi_{0.8}Mn_{1.2}Co_{0.8}O_{6.02}$ phase represents a state, which is very close to the spin-glass type phase whereas the other two phases are glassy FM in nature. As a confirmation to the magnetic behavior at low temperature and to correlate with the electronic conduction, the magnetic field dependent electrical resistivity measurements have been investigated throughout the temperature range.

**C. Electron transport properties**

Furthermore, we have carried out temperature dependent electrical resistivity ($\rho$) as well as isothermal magnetoresistance (*MR*) measurements for all samples. Figure 6 shows the electrical resistivity $\rho(T)$ in the presence and absence of the external magnetic field of 70 kOe. In the zero applied field condition, the resistivity increases rapidly as manganese is replaced by nickel or cobalt. It is noticed that the samples are insulating throughout the measured temperature range (80K ≤ *T* ≤ 400K) and the resistivity value is very high at low temperature, crossing instrument limitations. Such an insulating behavior was reported earlier by Zhao *et al.*[17] for the parent $La_{1.2}Bi_{0.8}Mn_2O_{6.20}$ compound, but the magnitude of the resistivity has been increased in the present compound by one order of magnitude i.e. ~$10^4 \Omega$ cm at 130K to be compared to ~$10^3 \Omega$ cm for the reported value.[17] However, the authors did not study the *MR* for this compound. We have studied $\rho(T)$ in an applied field of 70 kOe and we observed that the insulating behavior is robust down to low temperatures for all compounds. However, below the FM transition temperature, the $\rho(T)$ value decreases slightly for the $La_{1.2}Bi_{0.8}Mn_{1.2}Ni_{0.8}O_{5.88}$ and $La_{1.2}Bi_{0.8}Mn_{1.2}Co_{0.8}O_{6.02}$ compounds and the effect is significantly large for the parent compound. Thus, similarly to the $BiMnO_3$ none of these samples show an insulator-metal transition, even in the presence of high magnetic field. In the 80–400K range, the $\rho(T)$ curve confirms the insulating behavior of these materials (Fig. 6), all of them being FM below room temperature and the *MR* effect is present although small in magnitude. Hence, at low temperature there is a strong correlation between magnetic and electronic phase in these systems. To further characterize this correlation behavior we have investigated the isotherm *MR* measurements at low temperatures.

The magnetoresistance (*MR*) behavior of the three compounds shows a clear magnetic field dependent change in the resistivity at low temperature (Fig. 7). The *MR* value is calculated as *MR (%)* = [$\{\rho(7)-\rho(0)\}/\rho(0)$]x100, where $\rho(0)$ is the sample resistivity at 0 kOe and $\rho(7)$ under an applied field of ±70 kOe. It is observed that the obtained *MR* values are negative and consistent with the values reported for Bi-based perovskite.[17] The highest obtained negative *MR* value is around -67 % at 90K in an applied field of ±70 kOe for the parent $La_{1.2}Bi_{0.8}Mn_2O_{6.20}$ [Fig. 7(a)] and near $T_C$ the value is ~ -61%. Whereas, for $La_{1.2}Bi_{0.8}Mn_{1.2}Co_{0.8}O_{6.02}$ the isothermal *MR* value is only -17%, just below the $T_C$ (at 175K) and the highest value ~ -19% is obtained at 125K (the data was beyond the instrumental limit for $T \leq 120K$). The present data suggest that the *MR* values are irreversible in nature and almost constant in the FM region. However, in the case of $La_{1.2}Bi_{0.8}Mn_{1.2}Ni_{0.8}O_{5.88}$ the corresponding anisotropic

*MR* values are -12% (at 175K) and -20% (at 125K). The occurrence of anisotropic *MR* behavior at low temperatures similar to those in *M(H)* studies (Fig. 4) suggests the strongly correlated nature of field-induced magnetic and electronic behavior. Nevertheless, the obtained *MR* values are less than the expected values near the FM $T_C$. Moreover, for $La_{1.2}Bi_{0.8}Mn_2O_{6.20}$ the obtained *MR* value in the PM region [Fig. 7(a)] is around -25% (at 175K), which may be due to intergrain *MR* effect above $T_C$.[25] With increasing temperature the *MR* value gradually decreases and finally near the room temperature the obtained value is ~ -1% in ±70 kOe for all samples.

It is well known that in compounds with competing magnetic orders, a magnetic field favoring one kind of order in the spins also causes a large negative *MR*. Therefore, for the present systems this kind of field induced magnetotransport interactions are expected due to the presence of magnetic clusters. The obtained negative *MR* for all compounds (Fig. 7) can be well understood in the scenario of suppression of the electron scattering below the FM $T_C$ in the presence of an external field. Whereas near the room temperature, due to the thermal excitation the electron scattering mechanism is enhanced giving rise to lower *MR* values. The charge transport in this system turns out to be very sensitive to the FM ordering, and the magnetic fields readily induce a negative *MR* below $T_C$. Hence, all compounds show FM and insulating behavior (even at high magnetic field) with electronic phase separation at low temperature (FM and AFM clusters).

In order to throw light on the transport mechanisms operating in the present compounds, we have analyzed the low temperature behavior. Different models[26] have been proposed to describe the temperature dependent charge transport behavior for oxide system defined by $ln\ \rho \propto T^{-1/n}$, where *n* = 1, 2 or 4. Here, *n* = 1, corresponds to a simple thermal activation (TA) model, when *n* = 2, the hopping is referred as Efros-Shklovskii type hopping (ESH), controlled by Coulombic forces. When $n = 4$, there would be variable range hopping (VRH) and the hopping dynamics is controlled by collective excitation of the charge carriers. These models are compared to describe the zero field $\rho(T)$ behavior in our compounds. Figure 8(a) shows the TA model, $\ln \rho \propto T^{-1}$, which describes well the resistivity behavior above the FM transition $T_C$ with the activation energy, $E_a$, of 73, 82 and 86 meV for $La_{1.2}Bi_{0.8}Mn_2O_{6.20}$, $La_{1.2}Bi_{0.8}Mn_{1.2}Ni_{0.8}O_{5.88}$ and $La_{1.2}Bi_{0.8}Mn_{1.2}Co_{0.8}O_{6.02}$ respectively. This suggests that with increasing the doping concentration at Mn site the energy band gap increases gradually. This is consistent with the value reported by Chiba et al[10] for bismuth containing narrow band gap insulator. It is clearly noticed from Fig. 8(b) that, the ESH model, $\ln \rho \propto T^{-1/2}$, aptly describes the resistivity behavior for the doped compounds in the whole measured temperature range ($100K \leq T \leq 400K$) among the three classes of model generally applied to the perovskite materials.[27] The ESH type conductivity is usually observed when the coulomb interaction starts to play a key role in carriers hopping. The ESH or VRH models do not fit to the experimental data at high temperatures for the parent compound in contrast to the TA model. It must be emphasized that for Ni/Co doped samples, the resistivity evolution follows the VRH model only in the low temperature region ($T \leq 220K$) as shown in the inset of Fig. 8(b). This type of hopping conduction is typical to the phase separated systems (FM and AFM) where the charge carriers move by hopping between two localized electronic states. Then the conductivity results from hopping motion of the charge carriers through such localized states.[27]

**D. Magnetodielectric effects**

From the previous investigations it is clear that, among the three samples, the $La_{1.2}Bi_{0.8}Mn_{1.2}Co_{0.8}O_{6.02}$ is best for studying the dielectric properties as a function of external magnetic field and temperature, since it is an insulating ferromagnet with the highest $T_C$ (Fig. 2) as well as an energy band gap [Fig. 8(a)] compared to other phases. Therefore, the isothermal magnetodielectric studies were carried out on the $La_{1.2}Bi_{0.8}Mn_{1.2}Co_{0.8}O_{6.02}$ phase at different temperatures. Figure 9(a) shows the magnetic field dependent real ($\varepsilon'$) and imaginary ($\varepsilon''$) parts of the dielectric constant for $La_{1.2}Bi_{0.8}Mn_{1.2}Co_{0.8}O_{6.02}$ at 80K. A positive magneto-dielectric value of 0.25% is obtained around 80K (measured at 500kHz), which is calculated[28] as, $\Delta\varepsilon = [\{\varepsilon(H)-\varepsilon(0)\}/\varepsilon(0)] \times 100$, where $\varepsilon(0)$ is the sample capacitance at 0 kOe and $\varepsilon(H)$ under an applied field of ± 50kOe. The magneto-capacitance effect ($\Delta\varepsilon$) is found to be around 0.21%-0.25% in the temperature range of 10-80K respectively, quite similar to the behavior of $LaBiMn_{4/3}Co_{2/3}O_6$ thin film.[20] Above 90K, the $\Delta\varepsilon$ displays a gradual increase from 0.5% to 10% up to 300K. The positive value of $\Delta\varepsilon$ signifies the core dominated *MR* effect rather than other artifacts.[28] Hence, the obtained $\Delta\varepsilon$ value for $La_{1.2}Bi_{0.8}Mn_{1.2}Co_{0.8}O_{6.02}$ is higher near the magnetic ordering temperature ($T_C \sim$ 181K) and the low temperature values are comparable to the value of bulk $BiMnO_3$.[13] An intrinsic magnetodielectric effect arising from the spin lattice interaction present in the system could give rise to a magnetocapacitance variation, whereas, the reverse is not necessarily true always.[28] Hence, to further understand the origin of the magnetocapacitance effect, observed at low and high temperature range, the temperature dependent dielectric studies, in the presence and absence of magnetic field and the frequency dispersion of the dielectric constant measurements were carried out.

Figure 9(b), shows the temperature dependence of the real ($\varepsilon'$) and imaginary ($\varepsilon''$) parts of the dielectric constant of the sample, measured at 500kHz in the presence of magnetic fields. The obtained $\varepsilon'$ value varies from 30 to 3000 in the temperature

range of 10-300K. At low temperatures ($\leq$ 80K) both the $\varepsilon'$ and $\varepsilon''$ values are independent of frequency as well as temperature (not shown). Around 140-200K a rapid increase of frequency dependent dielectric constant behavior is observed. Similar kind of sudden increase in the dielectric constant behavior is reported in the literature for other systems.[29,30] The rapid increase in the dielectric constant values by two orders of magnitude around 150K at 500kHz [Fig. 9(b)] could arise from various factors like intrinsic to the system and/or extrinsic effects. In the case of intrinsic effect, the steep rise of the dielectric constant over a short range of temperature could arise around the onset of ferroelectric ordering in which the dipoles are subjected to a double well energy barrier. In that case, the dielectric constant obeys the Curie-Weiss behavior above the transition temperature, which is not the case in the present study (since the transition temperature is not well defined). In the case of extrinsic effects, the steep rise in the dielectric constant with temperature could be due to (i) barrier layer effects arising from the crystallographic twin boundaries (If any present in the system), (ii) Maxwell-Wagner kind of relaxation arising from the grain boundaries and (iii) barrier layer formation between the material and the metal contacts.[29-31] In the present case, the effect observed could be a combined effect of the grain boundary and at temperatures above 250K, the space charge effect arising due to the semiconducting behavior of the $La_{1.2}Bi_{0.8}Mn_{1.2}Co_{0.8}O_{6.02}$ sample associated with the contact effects. To simplify and understand the behavior of various microscopic effects influencing the dielectric studies, the knowledge about the distribution in the relaxation times of the extrinsic factors could be utilized. The relaxation time characteristic to the extrinsic components could be resolved in the frequency domain measurements at different temperatures.[29,30] Hence, observation of the frequency dispersion and its temperature dependent studies could throw more light on the effect observed.

Figure 10, shows the frequency dispersion (100Hz–1MHz) of the real ($\varepsilon'$) and imaginary ($\varepsilon''$) parts of dielectric constant for $La_{1.2}Bi_{0.8}Mn_{1.2}Co_{0.8}O_{6.02}$ at different temperatures. The frequency dependent $\varepsilon'$ exhibits a large value at low frequency (< 100Hz) and a step like decrease with increase in frequency (> 1kHz) at 130K. An associated Debye like relaxation peak in $\varepsilon''$ data is observed in the vicinity of the step like frequency dispersion observed in $\varepsilon'$ data. The observed relaxation peaks also shift towards higher frequencies with increasing temperature similar to an activation behavior. The frequency-dependent behavior of the real and imaginary part of the dielectric constant is commonly observed in the case of materials exhibiting colossal dielectric behavior ($\varepsilon'$ >1000) due to extrinsic effects.[29,30] Though the relaxation behavior looks like a Debye type relaxation (a characteristic of the intrinsic dipoles present in the system), the temperature dependent frequency shift suggests that the relaxation could be of Maxwell-Wagner type, which arises due to the presence of regions with different conductivities within the sample.[28-31] In the case of $La_{1.2}Bi_{0.8}Mn_{1.2}Co_{0.8}O_{6.02}$ the grain boundaries present and/or different magnetic clusters could act as a region of different conductivity as observed in the case of magnetic and electron transport studies. The large values in the low frequency $\varepsilon'$ imply that the grain boundary capacitance is larger than the bulk grain capacitance of the sample. Hence, the charge accumulated at the grain boundaries could give rise to a Maxwell-Wagner type relaxation phenomenon.[29,30] The gradual increase of the $\varepsilon'$ value at low frequency (for $T \geq 130K$) could arise due to the space charge (due to DC conduction) and other contact based effects. The detailed analysis of the influence of microstructure over the relaxation and the influence of different electrode materials with different sample thicknesses are beyond the scope of this article.

The capacitance measurements carried out at different temperatures and frequencies reveal the overlap of the grain boundary relaxation of the sample. Hence, it is important to select the suitable temperature and the frequency to understand the magneto-capacitance effect observed in these samples. It is observed from the electron transport properties that the resistivity increases exponentially (Fig. 6) with decreasing temperature, for $T < T_C$ carrier effects are not expected to play a relevant role in dielectric properties. Therefore, the capacitance measured at low temperature and high frequencies provides true intrinsic effects of the sample.[28] Based on the above mentioned observations one could clearly see that the magnetocapacitance effect ($\Delta\varepsilon$) is around 0.5 to 10% for $La_{1.2}Bi_{0.8}Mn_{1.2}Co_{0.8}O_{6.02}$ system and is dominated by the grain boundary conduction effect above 100K for all the measured frequency range (100Hz–1MHz). Moreover, the positive weak $\Delta\varepsilon$ effect of 0.21–0.25% observed at low temperatures ($T \leq 80K$), is expected to originate from the spin lattice interaction intrinsic to the system akin to $LaBiMn_{4/3}Co_{2/3}O_6$ thin film.[20] Note that, the studied temperature range is well below the onset of the ferromagnetic ordering of $La_{1.2}Bi_{0.8}Mn_{1.2}Co_{0.8}O_{6.02}$ sample with different types of magnetic clusters. In addition, the $6s^2$ lone pair electron present for $Bi^{3+}$ is expected to introduce an additional distortion, which in effect is coupled with the lattice. Hence, on varying the magnetic field, the alignment of magnetic clusters could distort the lattice, which in effect is reflected on the variation of the dielectric constant.[32]

## IV. CONCLUSIONS

The present investigations show that the substitution of nickel or cobalt for manganese in a bismuth based manganite such as $La_{1.2}Bi_{0.2}MnO_{6.20}$ not only allows

FM $T_C$ to be increased, but also enhances the insulating properties. Such an effect is of capital importance for the exploring spintronic materials since it allows significant magneto-dielectric properties to be reached. Although the physics of these materials is so far not completely understood it appears that three factors are at the origin of these particular properties. The first one concerns the coexistence of $Mn^{4+}$ and $Ni^{2+}$ or $Co^{2+}$ species in the same oxide matrix, which induce strong FM interactions in agreement with Goodenough-Kanamori rules, reinforcing in this way the FM $Mn^{3+}$-O-$Mn^{4+}$ interactions. The second factor deals with the existence of electronic phase separation where FM regions ($Mn^{4+}/Ni^{2+}$ or $Co^{2+}$ and $Mn^{3+}/Mn^{4+}$) compete with AFM regions ($Mn^{3+}/Mn^{3+}$ and $Mn^{3+}/Ni^{2+}$ or $Co^{2+}$), leading to large values of *MR* at low temperature. Finally, the third point concerns the spin-lattice interactions, connected to structural distortions introduced by the $Bi^{3+}$ *$6s^2$* lone pair, which are at the origin of the magnetodielectric effect (also pointed out by Montanari *et al.*[15] for centrosymmetric $BiMnO_3$). Further investigations are in progress in order to understand the exact nature of the interactions and the mechanism, which give rise to an intrinsic magnetodielectric effect in these oxides.

## Acknowledgements

We gratefully acknowledge Dr. W. Prellier and Prof. P. Nordblad, for scientific discussions and the STREP MaCoMuFi (NMP3-CT-2006-033221), the CNRS and the Ministry of Education and Research for financial support.

**Table. 1.** Lattice parameters, magnetic and electrical properties of $La_{1.2}Bi_{0.8}Mn_{2-x}(Ni/Co)_xO_6$. Where *a*, *b*, *c*, $R_b$ and $R_f$ are the lattice parameters, Bragg factor and fit factors, respectively. $T_C$ is the ferromagnetic Curie temperature, $\theta_p$ is the Curie-Weiss temperature, $\mu_{eff}$ is the effective paramagnetic moment, $M_S$ is the approximate saturation value (10K), $H_C$ is the coercive field (10K), $E_a$ is the activation energy, $\rho^{300}$ is the electrical resistivity (300K) and *MR(%)* is the maximum values of magnetoresistance.

| Compositions | $La_{1.2}Bi_{0.8}Mn_2O_{6.20}$ | $La_{1.2}Bi_{0.8}Mn_{1.2}Ni_{0.8}O_{5.88}$ | $La_{1.2}Bi_{0.8}Mn_{1.2}Co_{0.8}O_{6.02}$ |
|---|---|---|---|
| Space group | *Pnma* | *Pnma* | *Pnma* |
| *a* (Å) | 5.539(3) | 5.492(3) | 5.500(3) |
| *b* (Å) | 7.849(6) | 7.865(3) | 7.775(3) |
| *c* (Å) | 5.528(1) | 5.502(3) | 5.506(3) |
| Cell volume (Å³) | 240.38(2) | 234.64(4) | 235.48(2) |
| $R_b$(%) | 6.68 | 4.59 | 4.13 |
| $R_f$(%) | 10.1 | 8.04 | 7.56 |
| $T_C$ (K) | 103 | 178 | 181 |
| $\theta_p$ (K) | 140 | 200 | 205 |
| $\mu_{eff}$ ($\mu_B$/f.u.) | 7.67 | 5.66 | 6.70 |
| $M_S$ ($\mu_B$/f.u.) | 6.82 | 4.11 | 5.36 |
| $H_C$ (Oe) | 51 | 85 | 2721 |
| $\rho^{300}$ (Ω.cm) | 2 | 34 | 231 |
| $E_a$ (meV) | 73 | 82 | 86 |
| *MR(%)* | 67.5 | 20.5 | 19.5 |

**Figure captions**

**FIG. 1.** (Color online) Rietveld analysis of XRPD patterns of (a) $La_{1.2}Bi_{0.8}Mn_2O_{6.20}$ (b) $La_{1.2}Bi_{0.8}Mn_{1.2}Ni_{0.8}O_{5.88}$ and (c) $La_{1.2}Bi_{0.8}Mn_{1.2}Co_{0.8}O_{6.02}$ at room temperature. Open symbols are experimental data and the dotted, solid and vertical lines represent the calculated pattern, difference curve and Bragg position respectively.

**FIG. 2.** (Color online) Temperature dependent ZFC (open symbol) and FC (solid symbol) magnetization, M, for (a) $La_{1.2}Bi_{0.8}Mn_2O_{6.20}$ (square), $La_{1.2}Bi_{0.8}Mn_{1.2}Ni_{0.8}O_{5.88}$ (triangle) and $La_{1.2}Bi_{0.8}Mn_{1.2}Co_{0.8}O_{6.02}$ (circle) in an applied field of H = 1000 Oe, and (b) $La_{1.2}Bi_{0.8}Mn_{1.2}Co_{0.8}O_{6.02}$ (circle) in two different fields (H=100 Oe and 5000 Oe).

**FIG. 3.** (Color online) Temperature dependent inverse magnetic susceptibility, $\chi^{-1}$, plot for $La_{1.2}Bi_{0.8}Mn_2O_{6.20}$ (square), $La_{1.2}Bi_{0.8}Mn_{1.2}Ni_{0.8}O_{5.88}$ (triangle) and $La_{1.2}Bi_{0.8}Mn_{1.2}Co_{0.8}O_{6.02}$ (circle).

**FIG. 4.** (Color online) Field dependent isothermal magnetic hysteresis, M(H), curves at different temperatures for (a) $La_{1.2}Bi_{0.8}Mn_2O_{6.20}$ (b) $La_{1.2}Bi_{0.8}Mn_{1.2}Ni_{0.8}O_{5.88}$ and (c) $La_{1.2}Bi_{0.8}Mn_{1.2}Co_{0.8}O_{6.02}$. The inset figures show the enlarged version of M(H) at 10K.

**FIG. 5.** (Color online) Temperature dependent magnetic ac-susceptibility of (a) $La_{1.2}Bi_{0.8}Mn_2O_{6.20}$ (b) $La_{1.2}Bi_{0.8}Mn_{1.2}Ni_{0.8}O_{5.88}$ (in phase component, $\chi'$) and (c) $La_{1.2}Bi_{0.8}Mn_{1.2}Co_{0.8}O_{6.02}$ (both, in phase $\chi'$ and out of phase $\chi''$ components) at different frequencies ($h_{ac}$ = 10 Oe).

**FIG. 6.** (Color online) Temperature dependent electrical resistivity, $\rho$, of $La_{1.2}Bi_{0.8}Mn_2O_{6.20}$ (square) $La_{1.2}Bi_{0.8}Mn_{1.2}Ni_{0.8}O_{5.88}$ (triangle) and $La_{1.2}Bi_{0.8}Mn_{1.2}Co_{0.8}O_{6.02}$ (circle) in the presence (solid symbol) and absence (open symbol) of magnetic field (70 kOe).

**FIG. 7.** (Color online) The isothermal magnetoresistance MR(%) at different temperatures for (a) $La_{1.2}Bi_{0.8}Mn_2O_{6.20}$ (b) $La_{1.2}Bi_{0.8}Mn_{1.2}Ni_{0.8}O_{5.88}$ and (c) $La_{1.2}Bi_{0.8}Mn_{1.2}Co_{0.8}O_{6.02}$

**FIG. 8.** (Color online) Logarithm of resistivity versus $T^{-1/n}$ plots for $La_{1.2}Bi_{0.8}Mn_2O_{6.20}$ (square), $La_{1.2}Bi_{0.8}Mn_{1.2}Ni_{0.8}O_{5.88}$ (triangle) and $La_{1.2}Bi_{0.8}Mn_{1.2}Co_{0.8}O_{6.02}$ (circle): Open symbols and solid lines represent the experimental data and apparent fit to the different hopping models as described in the text.

**FIG. 9.** (Color online) (a) Magnetic field dependent isothermal real ($\varepsilon'$) and imaginary ($\varepsilon''$) parts of dielectric constants at 80K and (b) Temperature dependent $\varepsilon'$ and $\varepsilon''$ values in different magnetic fields for $La_{1.2}Bi_{0.8}Mn_{1.2}Co_{0.8}O_{6.02}$ at a frequency of 500 kHz.

**FIG. 10.** (Color online) Frequency dispersion behavior of the real ($\varepsilon'$) and imaginary ($\varepsilon''$) part of the dielectric constants at different temperatures for $La_{1.2}Bi_{0.8}Mn_{1.2}Co_{0.8}O_{6.02}$.

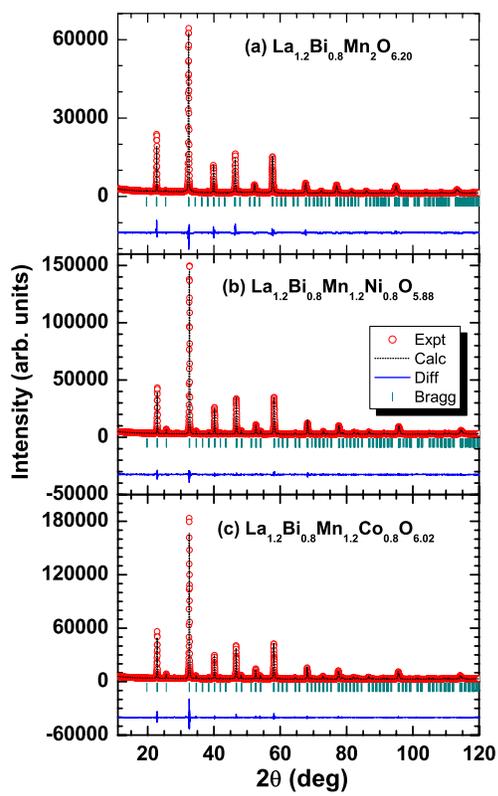

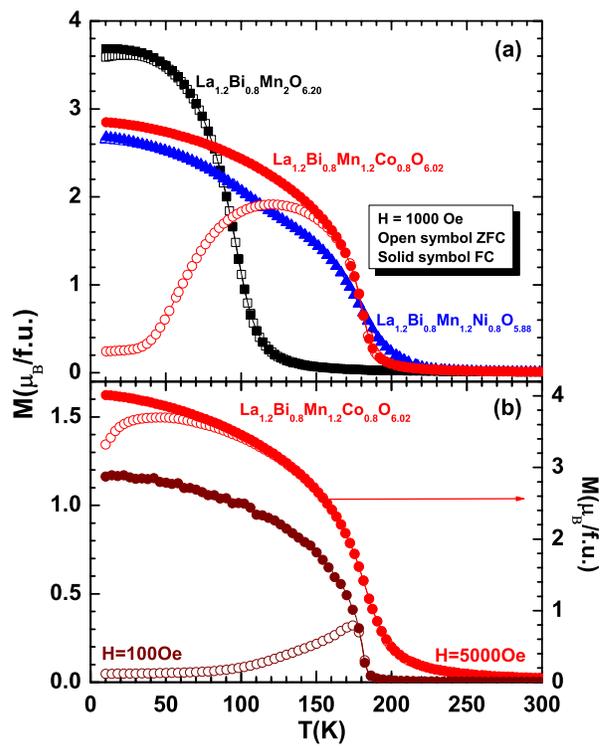

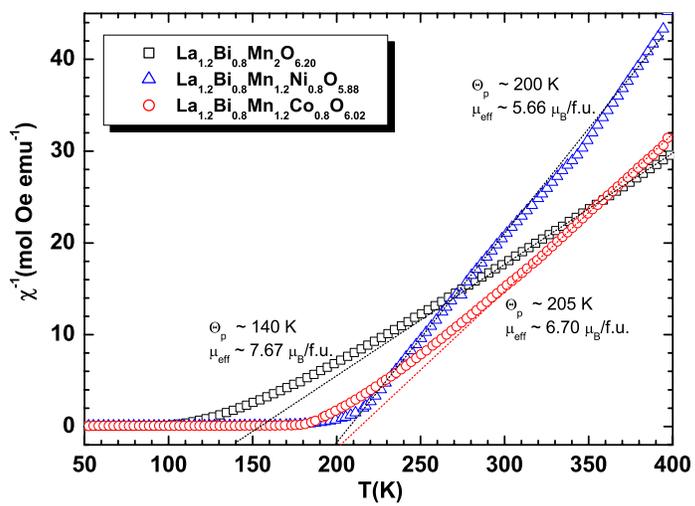

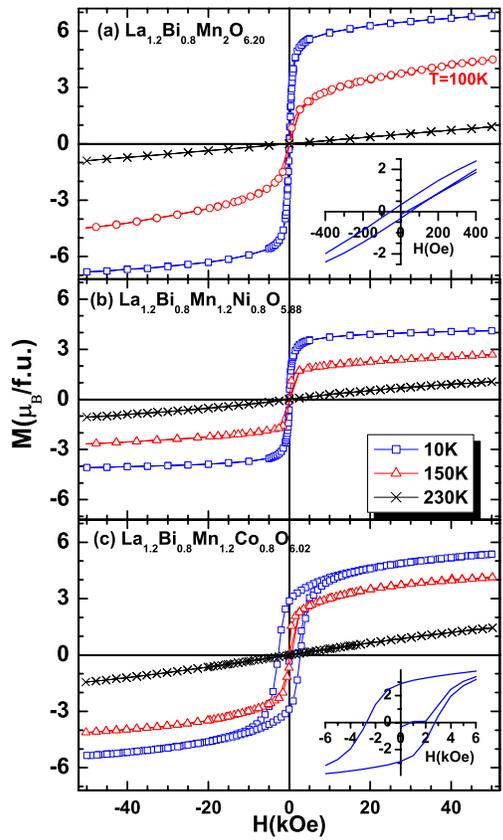

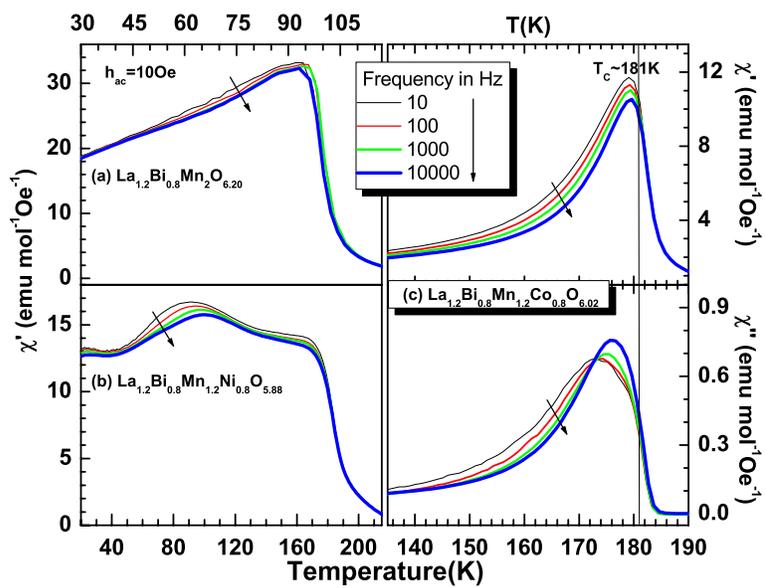

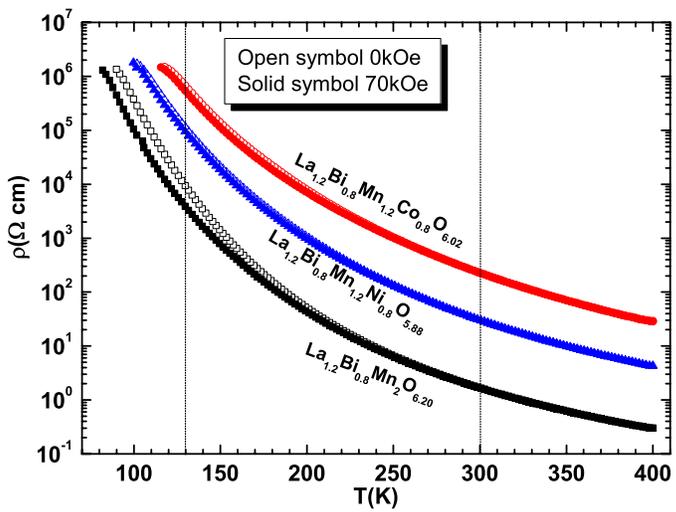

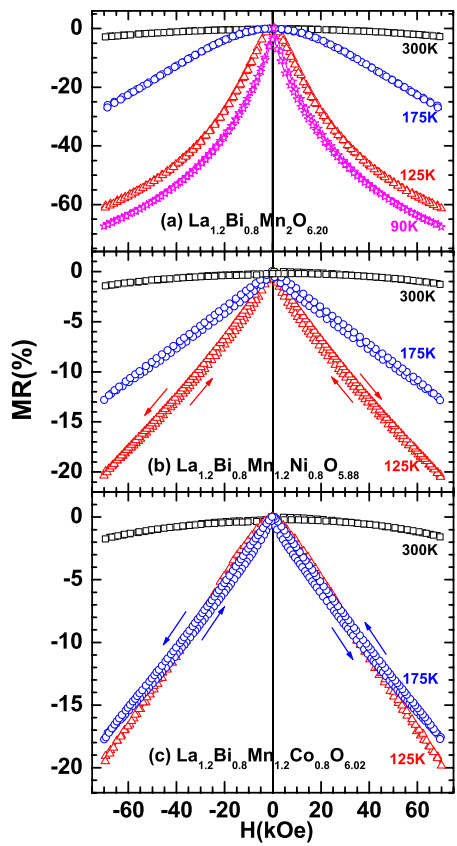

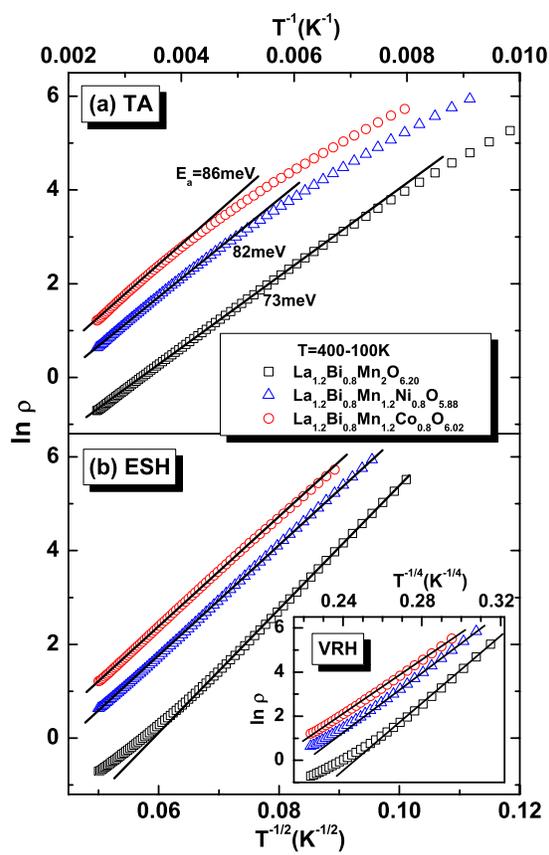

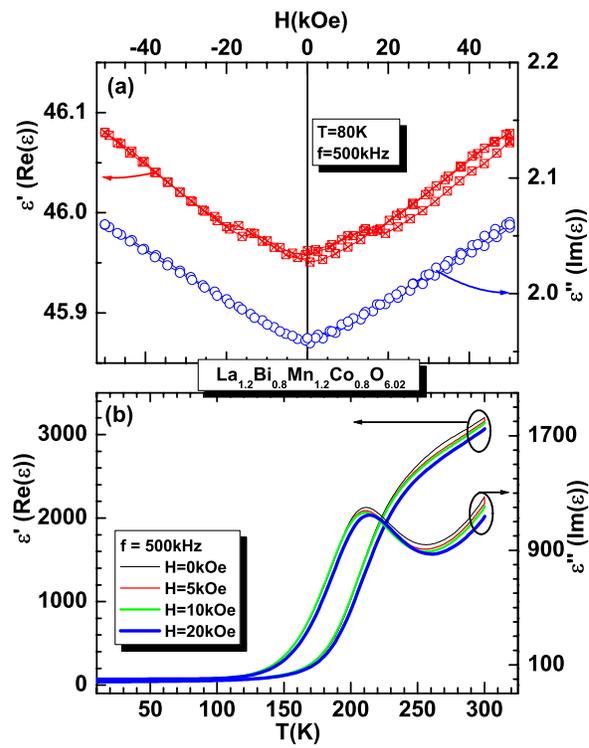

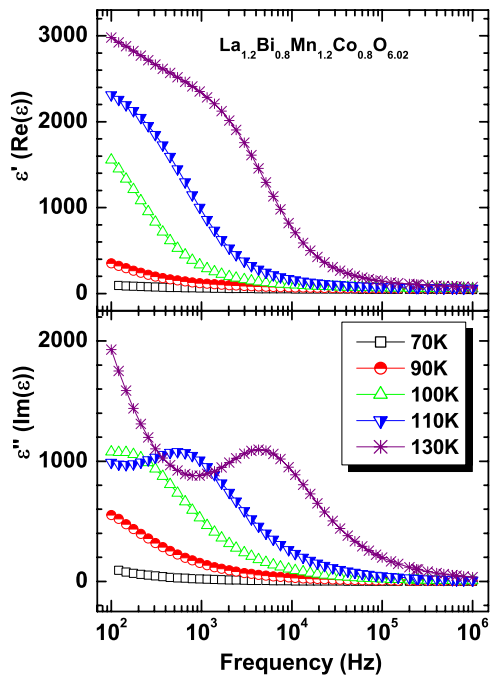